\documentclass[11pt,a4paper]{article}
\usepackage[margin=1in]{geometry}
\usepackage{graphicx}
\usepackage{amsmath}
\usepackage{amssymb}
\usepackage{cite}
\usepackage{url}
\usepackage{color}
\usepackage{xcolor}
\usepackage{subcaption}
\usepackage[most]{tcolorbox}
\usepackage{caption}

\newtcolorbox{csmbox}[2][]{
  colback=gray!5,
  colframe=blue!75!black,
  fonttitle=\bfseries,
  title=#2,
  #1
}

\definecolor{csmdarkblue}{RGB}{0,51,102}
\definecolor{csmlightblue}{RGB}{102,153,204}

\begin{document}

\title{\textbf{Multicellular Feedback Control Strategies in Synthetic Microbial Consortia: From Embedded to Distributed Control}}

\author{Mario di Bernardo\\
Department of Electrical Engineering and Information Technology\\
University of Naples Federico II, 80125 Naples, Italy\\
Scuola Superiore Meridionale, Naples, Italy\\
\texttt{mario.dibernardo@unina.it}}

\date{}

\maketitle

\begin{abstract}
Living organisms rely on endogenous feedback mechanisms to maintain homeostasis in the presence of uncertainty and environmental fluctuations. An emerging challenge at the interface of control systems engineering and synthetic biology is the design of reliable feedback strategies to regulate cellular behavior and collective biological functions. In this article, we review recent advances in multicellular feedback control, where sensing, computation, and actuation are distributed across different cell populations within synthetic microbial consortia, giving rise to biological multiagent control systems governed by molecular communication.
From a control-theoretic perspective, these consortia can be interpreted as distributed biomolecular control systems, where coordination among populations replace embedded regulation. We survey theoretical frameworks, control architectures, and modeling approaches, ranging from aggregate population-level dynamics to spatially aware agent-based simulations, and discuss experimental demonstrations in engineered \textit{Escherichia coli} consortia. We highlight how distributing control functions across populations can reduce metabolic burden, mitigate retroactivity, improve robustness to uncertainty, and enable modular reuse of control components.
Beyond regulation of gene expression, we discuss the emerging problem of population composition control, where coordination among growing and competing cell populations becomes an integral part of the control objective. Finally, we outline key open challenges that must be addressed before multicellular 
control strategies can be deployed in real-world applications such as biomanufacturing, 
environmental remediation, and therapeutic systems. These challenges span modeling 
and simulation, experimental platform development, coordination and composition 
control, and long-term evolutionary stability.
\end{abstract}


\section{Introduction}
\label{sec:intro}
The engineering of living systems represents one of the most compelling frontiers where control theory meets biological implementation \cite{delvecchio2014bfs,delvecchio2016control,hsiao2018control,khammash2022cybergenetics}. While traditional control applications often benefit from well-characterized physical 
models, biological systems present distinct challenges as they are inherently stochastic, 
operate at molecular scales where deterministic approximations become questionable, 
grow and evolve over time, and must function robustly despite parameter uncertainty 
and environmental perturbations that substantially exceed typical engineering scenarios. 
These characteristics demand both new theoretical frameworks and novel implementation 
strategies that extend classical control paradigms while respecting the constraints 
imposed by living matter.

Yet these same challenges reveal a profound opportunity. Living organisms themselves are a proof that robust control at molecular scales is achievable, as they can maintain homeostasis through sophisticated feedback mechanisms that regulate internal states despite perturbations and uncertainties. From hormone secretion and signaling pathways in multicellular organisms to bacterial chemotaxis, naturally evolved negative feedback loops accomplish what appears daunting from an engineering perspective achieving reliable regulation amid extreme stochasticity and parameter variation. This natural precedent has inspired synthetic biologists to engineer control systems within living cells \cite{cameron2014brief,filo2023biomolecular,khammash2022cybergenetics,ruolo2021control}, opening new possibilities for applications ranging from optimized bioproduction \cite{sosa2023maximizing} to targeted therapeutic delivery \cite{din2016synchronized}.The central challenge lies in translating natural regulatory principles into 
rationally designed implementations that are both experimentally tractable and 
suitable for deployment in applications.

Control strategies for synthetically engineered biological systems go from fully external to fully embedded implementations. External control employs computer algorithms interfaced with cells through sensors (e.g. fluorescence microscopy) and actuators (e.g. chemical inducers, optogenetics), providing excellent performance and flexibility for laboratory applications \cite{milias2016automated,uhlendorf2012long,menolascina2014vivo}. However, this approach requires specialized hardware, continuous monitoring, and precise environmental control, making it unsuitable for autonomous applications such as \emph{in vivo} therapeutics or industrial bioreactors where precise real-time monitoring and external interventions are impractical \cite{perrino2021control}. 

Embedded biomolecular controllers implement complete feedback loops within cells using synthetic gene regulatory networks, achieving autonomous operation without external hardware. Notable achievements include antithetic integral feedback controllers, now a common motif in the field, that guarantee robust perfect adaptation through molecular sequestration \cite{briat2016antithetic,aoki2019universal}, biomolecular PID controllers implementing proportional, integral, and derivative actions \cite{filo2022hierarchy,chevalier2019design}, and various circuits demonstrating precise gene expression regulation in bacteria and mammalian cells \cite{frei2022genetic,samaniego2021ultrasensitive}. 

While single-cell embedded controllers have demonstrated impressive capabilities, they face three fundamental architectural limitations arising from centralizing all control functions within individual cells. First, metabolic burden emerges when synthetic circuits consume cellular resources (ribosomes, RNA polymerase, ATP, and metabolic intermediates) competing with endogenous processes \cite{mcbride2021predicting,ceroni2018burden}. This burden scales with circuit complexity, slowing growth and creating selective pressure for loss-of-function mutations that disrupt the controller but confer growth advantages. Second, retroactivity occurs when biological modules are connected, with downstream components affecting upstream behavior through loading effects such as ribosome or polymerase competition \cite{del2008modular,ceroni2015burden}. This violates the modularity assumption essential for plug-and-play design, making it difficult to predict assembled circuit performance from the characterisation of its individual components. Third, poor modularity emerges because once cells are engineered with embedded controllers, any change in control strategy requires a complete system re-engineering, limiting adaptability and reuse of existing designs. 
More broadly, there is currently no systematic methodology to translate abstract control laws into implementable biochemical reactions, making biomolecular controller design a largely ad-hoc process regardless of the chosen architecture.

To address both the architectural and ecological limitations of single-cell implementations a promising strategy is distributing regulatory and control functionalities across distinct cell populations within microbial consortia. In~\cite{fiore2017silico}, we introduced the concept of \emph{multicellular feedback 
control}, in which sensing, computation, and actuation are distributed across 
dedicated populations. This approach 
treats synthetic consortia as biological multiagent systems, a perspective we 
develop formally in Section~\ref{sec:multiagent}, where control functions are 
distributed across populations rather than concentrated within individual cells. 
Populations communicate via diffusible molecular signals and collectively implement 
closed-loop regulation of a target process. This architecture reduces metabolic 
burden per cell, mitigates retroactivity between controller and plant components, 
and enables modular reuse of controller populations across different 
applications~\cite{fiore2017silico,martinelli2022multicellular,martinelli2025multicellular}.

However, distributing control across populations introduces orchestration 
challenges absent in single-cell implementations. How can appropriate population 
compositions be maintained over time despite differential growth rates arising 
from metabolic burden and resource 
competition~\cite{Balagadde2008,Scott2017,Stephens2019,Fedorec2021,Fiore2021Chemostat,Brancato2024DualChamber,gutierrez2022dynamic,grandel2025long}? 
How can communication protocols based on quorum sensing remain effective in 
the presence of spatial gradients, diffusion delays, and signal 
degradation~\cite{boo2021quorum}? What control architectures can ensure reliable 
collective behaviour despite cell-to-cell variability and the absence of explicit 
error correction in molecular signalling~\cite{danino2010synchronized,chen2015emergent}?

These challenges arise because microbial consortia constitute a distinct class 
of multiagent systems, characterised by stochastic dynamics, continuous population 
growth, and communication through analog molecular diffusion. These properties 
differ substantially from traditional multiagent systems in robotics or distributed 
computing~\cite{grandel2021control}. Addressing them requires control strategies 
that operate across molecular, cellular, and population scales simultaneously.

This review synthesizes theoretical foundations, design principles, computational modeling approaches, and experimental implementation strategies for multicellular feedback control in synthetic bacterial consortia, with emphasis on coordination challenges unique to biological multiagent systems. We examine how distributing sensing, computation, and actuation functions across populations addresses architectural limitations while introducing new orchestration problems requiring explicit management. Drawing extensively on recent computational \cite{fiore2017silico} and experimental \cite{salzano2025vivo} demonstrations as illustrative case studies, we identify critical open challenges in evolutionary stability, scalability, and biological realizability that must be addressed to translate laboratory demonstrations into industrial and therapeutic applications. The review is organized as follows. Section~\ref{sec:multiagent} presents design 
principles for distributed multicellular architectures, including quorum sensing 
communication channels and the fundamental two-population controller framework, 
with explicit comparison to traditional multiagent systems. 
Section~\ref{sec:validation} examines computational validation approaches spanning 
aggregate population models and agent-based simulations, followed by experimental 
validation in engineered \textit{E.~coli} consortia. Section~\ref{sec:pid} extends 
the basic architecture to the multicellular PID controller family. 
Section~\ref{sec:composition} addresses the challenge of maintaining stable 
population ratios. Finally, Section~\ref{sec:conclusions} identifies outstanding 
research challenges and concludes with perspectives on future directions.

\section{Multicellular Control Architectures}
\label{sec:multiagent}
\begin{figure}[tbp]
\centering
\includegraphics[width=1\linewidth]{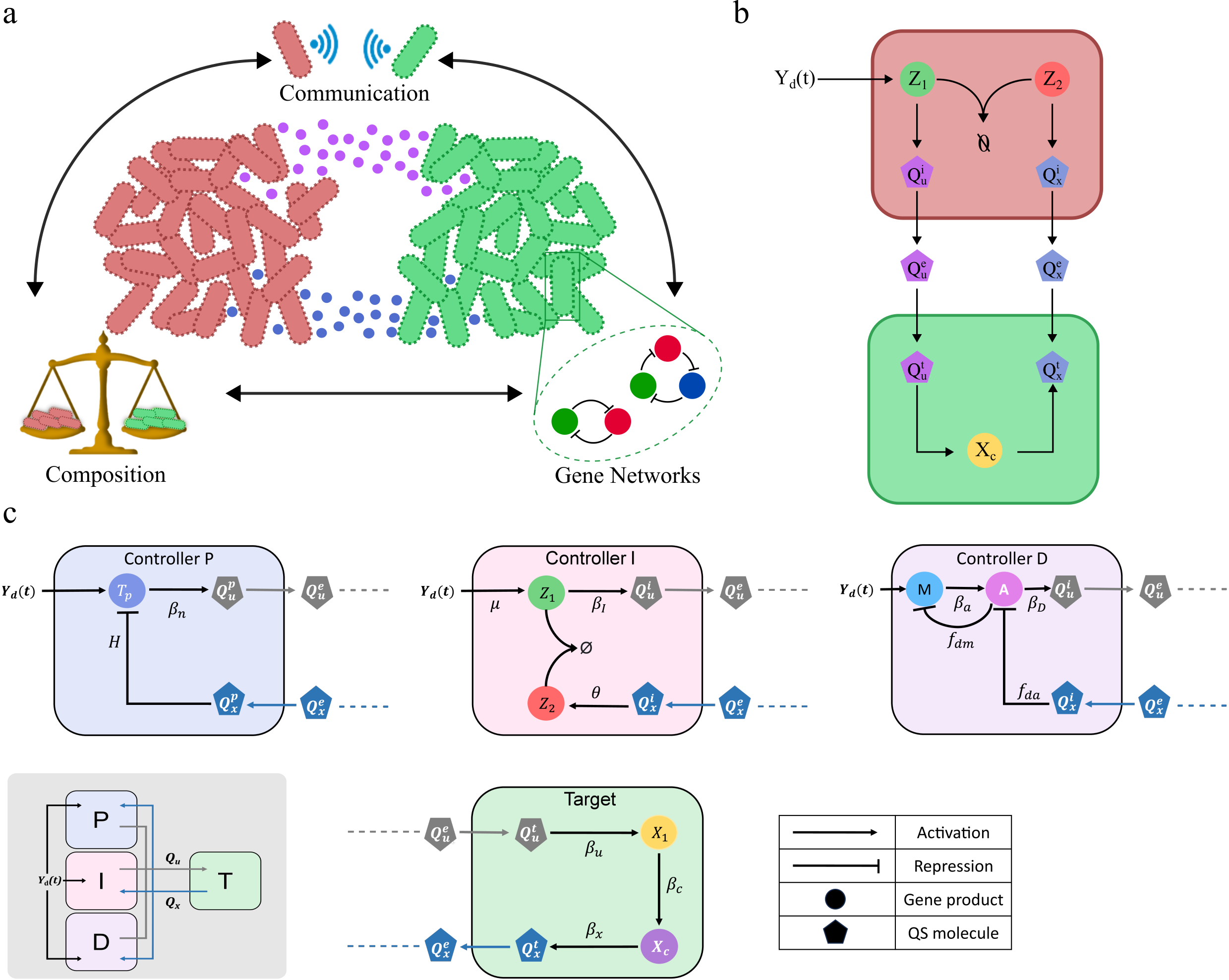}
\caption{\footnotesize\textbf{Multicellular control conceptual framework.}
(A) Key ingredients of a cellular consortium.
(B) Two-population architecture in which a \emph{Controller} cell population regulates the expression of a gene within a \emph{Target} population through quorum sensing (QS)--mediated communication, forming a closed feedback loop (reproduced from \cite{martinelli2022multicellular}).
(C) Four-population architecture implementing a distributed PID control, with separate cell populations responsible for proportional, integral, and derivative actions whose combined signals regulate the target population, enabling robust and tunable control of collective behavior (reproduced from \cite{martinelli2025multicellular}).}
\label{fig:multicellular_control_framework}
\end{figure}

The design of a multicellular feedback control system requires engineering three interdependent components (see Fig.~\ref{fig:multicellular_control_framework}a): the gene regulatory networks (GRNs) embedded within each population encoding the desired functions, the communication channels that enable information exchange between populations, and the consortium composition that determines the relative abundance of each population. 

The GRNs define the internal dynamics of each cell type, encoding sensing, 
computation, or actuation functions depending on the population's role in the control architecture. In bacterial systems, communication between populations is typically mediated by diffusible quorum sensing (QS) molecules, which act 
as molecular signals carrying information about cellular states across the consortium. Finally, the composition of the consortium, that is, the ratio of controller to target cells, directly affects closed-loop performance, as it determines the effectiveness of both the actuation and feedback pathways.

These three components are tightly coupled. The GRNs produce and respond to QS signals, the communication dynamics depend on population densities, and the composition evolves over time due to differential growth rates that may themselves depend on the metabolic burden imposed by the synthetic circuits. A successful multicellular control design must therefore address all three aspects in an integrated manner. In the following, we first present the two-population architecture that forms the basis of multicellular feedback control, then examine the quorum sensing systems that provide the communication substrate, and finally derive the mathematical framework describing the coupled dynamics of the consortium.

\subsection{Two-Population control architecture}
\label{sec:two_pops}

The simplest multicellular control architecture consists of two interacting populations, {\em controllers} and {\em targets}, which collectively implement closed-loop regulation of a biological process of interest \cite{fiore2017silico,salzano2025vivo}. This architecture, illustrated in Fig.~\ref{fig:multicellular_control_framework}b, establishes the basic framework upon which more sophisticated control strategies can be built.

Controller cells perform three key functions in the distributed control loop: they sense the output of target cells, compute the error between a desired reference and the measured output, and produce an actuation signal proportional to the computed control action. The reference signal is typically provided as an external inducer, while sensing and actuation are mediated by orthogonal quorum sensing (QS) channels, described in detail in Section~\ref{sec:qs}. An error computation mechanism is implemented through molecular mechanisms such as competing transcriptional regulators or molecular titration circuits as first proposed in \cite{annunziata2017orthogonal} and discussed in \cite{cuba2021ultrasensitive}.

Target cells execute the desired biological function, representing the plant in control terminology, while simultaneously reporting their state back to the controllers. The target genetic circuit responds to the control signal by modulating expression of the gene or pathway of interest. Concurrently, target cells synthesize a feedback signaling molecule whose production rate reflects the current state of the controlled output, thereby closing the feedback loop.

This architecture directly addresses the limitations of single-cell embedded control discussed in Section~\ref{sec:intro}. Metabolic burden is reduced because each population carries only a subset of the full circuit. Retroactivity between controller and plant is mitigated since their molecular components never interact within the same cellular compartment. Modularity is enhanced because the same controller population can, in principle, regulate different target populations by establishing appropriate communication channels. The realization of this architecture requires a reliable molecular communication system capable of transmitting information bidirectionally between populations.

\subsection{Quorum Sensing as Communication Channels}
\label{sec:qs}

Cell-to-cell communication through quorum sensing (QS) provides the molecular substrate enabling distributed control across populations. This naturally evolved bacterial communication system consists of matched pairs of signaling molecules and transcriptional regulators, and has been extensively characterized for synthetic biology applications \cite{boo2021quorum}. Sender cells produce diffusible signaling molecules, such as acyl-homoserine lactones (AHLs), that cross cell membranes into the extracellular environment following a concentration gradient. Receiver cells express cognate receptors that bind these molecules and modulate transcription of target genes, creating a communication channel between spatially separated cells.

The dynamics of QS-mediated communication can be captured through coupled differential 
equations describing intracellular production, cross-membrane exchange, and extracellular 
diffusion~\cite{fiore2017silico}. We present here an aggregate population model in which 
each population is treated as homogeneous, represented by a single effective cell whose 
state variables correspond to population-averaged concentrations. For a signalling molecule 
in sender cells, the intracellular concentration $Q^s$ evolves as
\begin{equation}
\label{eq:qs_sender}
\frac{dQ^s}{dt} = f_{\text{prod}} - \gamma_Q Q^s - \eta_s(Q^s - Q_e),
\end{equation}
where $f_{\text{prod}}$ denotes the production rate, $\gamma_Q$ is the intracellular 
dilution and degradation rate, and the final term captures diffusion across the cell 
membrane with rate $\eta_s$. Here $Q_e$ represents the extracellular concentration of the 
QS molecule. In receiver cells, which lack production capability, the intracellular 
dynamics reduce to
\begin{equation}
\label{eq:qs_receiver}
\frac{dQ^r}{dt} = -\gamma_Q Q^r - \eta_r(Q^r - Q_e),
\end{equation}
where for simplicity we assume the same dilution and degradation rate in both populations.

The extracellular concentration $Q_e(\mathbf{x},t)$, which in general depends on both spatial position $\mathbf{x}$ and time $t$, evolves according to:
\begin{equation}
\label{eq:qs_extracellular}
\frac{\partial Q_e}{\partial t} = \eta_s(Q^s_i - Q_e)N_s + \eta_r(Q^r_i - Q_e)N_r - \gamma_e Q_e + \Theta \nabla^2 Q_e
\end{equation}
where $N_s$ and $N_r$ are the densities of sender and receiver populations respectively, $\gamma_e$ is the extracellular degradation rate, and $\Theta$ is the spatial diffusion coefficient. The term $\Theta \nabla^2 Q_e$ accounts for spatial diffusion in structured environments or under limited mixing; it can be neglected when diffusion is fast relative to other dynamics, as in well-mixed microfluidic chambers or bioreactors.

These equations reveal key design considerations for multicellular control. Membrane exchange rates $\eta$ must be sufficiently fast relative to cellular dynamics to enable timely communication, as slow diffusion introduces delays that might degrade control performance. The extracellular degradation rate $\gamma_e$ determines how rapidly signals decay in the environment. Fast degradation can limit the spatial range over which populations can coordinate but improves temporal resolution by preventing signal accumulation. In structured environments such as biofilms or microcolonies, the diffusion coefficient $\Theta$ becomes critical, as spatial gradients can create pronounced concentration differences that disrupt coordination between populations \cite{tamsir2011robust},\cite{kim2019long}.

Several properties of QS systems are particularly relevant for their use in feedback control architectures. First, the availability of orthogonal systems that operate in parallel without crosstalk enables multiple independent communication channels within the same consortium. Well-characterized orthogonal pairs include the Lux system (3-oxo-C6-HSL) and Las system (3-oxo-C12-HSL), which exhibit minimal cross-activation when properly configured \cite{kylilis2018tools}. Second, QS systems are tunable. Production rates can be adjusted through promoter strength and ribosome binding site engineering, while degradation rates can be modulated by adding degradation tags to the QS synthase proteins. Third, QS signaling is inherently analog, with output levels varying continuously as a function of signal concentration according to Hill-type dose-response curves. However, the effective dynamic range is constrained by saturation of regulatory proteins at high inducer concentrations and by basal expression at low concentrations, limiting the amplitude available for control. These characteristics, orthogonality, tunability, and analog response with bounded dynamic range, define the design space within which multicellular control architectures must operate.

\subsection{Mathematical Model}
\label{sec:math_model}

We now derive a mathematical description of the two-population architecture 
illustrated in Fig.~\ref{fig:multicellular_control_framework}b. The model 
captures the dynamics of three coupled subsystems: the gene regulatory network 
within controller cells, the bidirectional QS communication channels, and the 
gene regulatory network within target cells. We adopt an aggregate population 
formulation in which each population is treated as homogeneous and well-mixed; 
extensions to spatially explicit agent-based models are discussed in 
Section~\ref{sec:validation}.

\subsubsection{Controller Population Dynamics}
The controller population implements error computation through an antithetic 
motif~\cite{briat2016antithetic}, in which two molecular species $Z_1$ and $Z_2$ 
mutually sequester one another. Species $Z_1$ is produced at a rate proportional 
to the reference signal $Y_d(t)$, provided as an external inducer, while species 
$Z_2$ is produced in response to the feedback QS molecule $Q_x$ received from the 
target population. The sequestration reaction $Z_1 + Z_2 \to \emptyset$ implements 
a comparison between reference and feedback signals: the net concentration of free 
$Z_1$ encodes the control error. The dynamics are given by
\begin{align}
\frac{dZ_1}{dt} &= \mu Y_d - \gamma_z Z_1 Z_2 - \gamma Z_1 \label{eq:Z1}\\
\frac{dZ_2}{dt} &= \theta Q_x^i - \gamma_z Z_1 Z_2 - \gamma Z_2 \label{eq:Z2}
\end{align}
where $\mu$ is the production rate constant for $Z_1$, $\theta$ governs the 
activation of $Z_2$ by the intracellular feedback signal $Q_x^i$, $\gamma_z$ is 
the sequestration rate, and $\gamma$ is the dilution rate due to cell growth.

The control signal is encoded in the QS molecule $Q_u$, whose intracellular 
concentration $Q_u^i$ in controller cells evolves as
\begin{equation}
\frac{dQ_u^i}{dt} = \beta_u Z_1 - \gamma Q_u^i + \eta(Q_u^e - Q_u^i)
\label{eq:Qu_controller}
\end{equation}
where $\beta_u$ is the production rate proportional to the free $Z_1$ concentration, 
and $\eta$ is the membrane diffusion rate governing exchange with the extracellular 
pool $Q_u^e$. For simplicity, we assume a common dilution rate $\gamma$ for all 
intracellular species; this assumption can be relaxed when species-specific 
degradation is significant.

\subsubsection{Target Population Dynamics}

Target cells respond to the control signal $Q_u$ by modulating expression of the 
controlled output $X_c$, and simultaneously produce the feedback signal $Q_x$ to 
report their state to the controllers. The controlled species evolves according to
\begin{equation}
\frac{dX_c}{dt} = f(Q_u^t) - \gamma X_c
\label{eq:Xc}
\end{equation}
where $Q_u^t$ denotes the intracellular concentration of the control signal in 
target cells and $f(\cdot)$ is a Hill-type activation function
\begin{equation}
f(Q_u^t) = \alpha_0 + \frac{\alpha_{\max} - \alpha_0}{1 + (K_u / Q_u^t)^{n_u}}
\label{eq:hill}
\end{equation}
with basal expression $\alpha_0$, maximal expression $\alpha_{\max}$, dissociation 
constant $K_u$, and Hill coefficient $n_u$.

The feedback QS molecule $Q_x$ is produced in proportion to the controlled output 
$X_c$. Its intracellular concentration in target cells evolves as
\begin{equation}
\frac{dQ_x^t}{dt} = \beta_x X_c - \gamma Q_x^t + \eta(Q_x^e - Q_x^t)
\label{eq:Qx_target}
\end{equation}
where $\beta_x$ is the production rate constant.

\subsubsection{Quorum Sensing Communication Channels}

The two populations exchange information through two orthogonal QS channels: the 
control channel carrying $Q_u$ from controllers to targets, and the feedback channel 
carrying $Q_x$ from targets to controllers. In each case, the intracellular 
concentration in receiver cells (which lack production capability) equilibrates 
with the extracellular pool through passive diffusion.

For the control signal in target cells:
\begin{equation}
\frac{dQ_u^t}{dt} = -\gamma Q_u^t + \eta(Q_u^e - Q_u^t)
\label{eq:Qu_target}
\end{equation}

For the feedback signal in controller cells:
\begin{equation}
\frac{dQ_x^i}{dt} = -\gamma Q_x^i + \eta(Q_x^e - Q_x^i)
\label{eq:Qx_controller}
\end{equation}

The extracellular concentrations depend on the balance between production by sender 
cells, uptake by receiver cells, and degradation in the environment. Assuming a 
well-mixed regime where spatial diffusion is fast relative to other dynamics, the 
extracellular concentration of the control signal evolves as
\begin{equation}
\frac{dQ_u^e}{dt} = N_c \eta (Q_u^i - Q_u^e) + N_t \eta (Q_u^t - Q_u^e) - \gamma_e Q_u^e
\label{eq:Qu_ext}
\end{equation}
where $N_c$ and $N_t$ are the densities of controller and target populations 
respectively, and $\gamma_e$ is the extracellular degradation rate. Similarly, 
for the feedback signal:
\begin{equation}
\frac{dQ_x^e}{dt} = N_t \eta (Q_x^t - Q_x^e) + N_c \eta (Q_x^i - Q_x^e) - \gamma_e Q_x^e
\label{eq:Qx_ext}
\end{equation}

\subsubsection{Closed-Loop System}

Equations~\eqref{eq:Z1}--\eqref{eq:Qx_ext} constitute a coupled dynamical system 
describing the closed-loop behaviour of the multicellular controller. The reference 
signal $Y_d(t)$ acts as the external input, the controlled output $X_c$ (or 
equivalently the feedback signal $Q_x$) represents the regulated variable, and the 
population densities $N_c$ and $N_t$ enter as parameters that affect the effective 
gains of both actuation and feedback pathways.

Under appropriate conditions, this architecture implements an integral feedback 
controller. Specifically, when the sequestration rate $\gamma_z$ is sufficiently 
high that the sequestration flux dominates dilution (i.e., $\gamma_z Z_1 Z_2 \gg 
\gamma Z_i$ for $i = 1,2$), the system achieves $Z_1 \approx Z_2$ at steady state, 
which implies $\mu Y_d \approx \theta Q_x^i$. This enforces a fixed relationship 
between reference and output that is largely insensitive to parameter variations 
in the target population~\cite{fiore2017silico}. This property, known as robust 
perfect adaptation, is central to the regulatory performance of the multicellular 
architecture, which implements a distributed version of the embedded antithetic 
feedback control strategy proposed in~\cite{briat2016design}. When dilution is not 
negligible relative to sequestration, the integrator becomes ``leaky'' and perfect 
adaptation is compromised; achieving the required separation of timescales is 
therefore an important consideration in the design of both single-cell and 
multicellular antithetic controllers~\cite{qian2018realizing}.

\subsubsection{Implementation Considerations}

Implementing multicellular feedback control introduces tuning challenges that go beyond those encountered in single-cell synthetic circuits, as parameters in physically separated populations must be coordinated to achieve desired closed-loop behavior.

A first consideration concerns the effective loop gain, which in the multicellular architecture depends on parameters distributed across both populations and the communication channels connecting them. In a single-cell controller, gain tuning involves adjusting promoter strengths and ribosome binding sites within one genetic context. In the multicellular setting, the overall gain is the product of contributions from controller gene expression, QS signal production and transmission, and target response characteristics. This distributed nature complicates systematic tuning: modifying a promoter in the controller population affects only part of the loop, and the resulting change in closed-loop behavior depends on the (potentially uncertain) characteristics of the target population and communication channels. Libraries of characterized genetic parts \cite{annunziata2017orthogonal} can facilitate exploration of this design space, but predicting closed-loop performance from individual component characterization remains challenging.

A second consideration is the population ratio. Unlike parameters internal to cells, the relative abundance of controllers and targets is influenced by growth dynamics and may drift over time. Experimental studies have demonstrated that functional regulation is maintained across controller-to-target ratios ranging from approximately 1:5 to 5:1 \cite{salzano2025vivo}, providing substantial operating margin. This robustness arises because the feedback mechanism partially compensates for composition changes: an increase in controller abundance strengthens actuation but also increases consumption of feedback signal, attenuating the net effect on closed-loop gain. Nevertheless, this self-compensation has limits, and severe composition imbalances degrade performance; a challenge we address in Section~\ref{sec:composition}.

Finally, the communication channels impose constraints absent in single-cell implementations. The dynamic range of QS-mediated signaling is bounded by promoter leakiness at low signal concentrations and receptor saturation at high concentrations, limiting the effective operating region of the controller. Diffusion delays between populations introduce phase lag that can destabilize the feedback loop if gain is set too high. These constraints couple the achievable bandwidth and stability margins to physical parameters of the growth environment, including cell density, mixing regime, and chamber geometry in microfluidic implementations.

\subsection{Multicellular Consortia as Biological Multiagent Systems}

Multicellular bacterial consortia represent a unique class of multiagent systems that differ fundamentally from traditional engineered multiagent systems in robotics, distributed computing, or cooperative control. These differences, summarized in Table~\ref{tab:multiagent_comparison}, highlight both challenges and opportunities specific to biological implementations, necessitating novel control design approaches that account for the unique physics and biology of microbial systems.

\begin{table}[ht]
\centering
\caption{Comparison of biological and traditional multiagent systems}
\label{tab:multiagent_comparison}
\begin{tabular}{p{3.5cm}p{5cm}p{5cm}}
\hline
\textbf{Property} & \textbf{Traditional Systems} & \textbf{Biological Consortia} \\
\hline
Agent dynamics & Deterministic or near-deterministic & Stochastic (molecular noise) \\
Agent population & Fixed or controllable & Dynamic (growth and division)  \\
Communication & Digital, error-corrected, high-bandwidth & Analog molecular diffusion, no error correction \\
Communication timescale & Milliseconds to seconds & Minutes to hours \\
Agent identity & Permanent, fixed function & Can change via differentiation or stochastic switching \\
Parameter uncertainty & Typically single-digit percentages & Significant (20-30\% or more across cells) \\
Control bandwidth & High (millisecond response) & Limited (minute-to-hour response) \\
Failure modes & Hardware/software faults (rare, detectable) & Mutations (continuous, often undetectable)\\
Evolutionary pressure & None & Continuous selection against metabolic burden \\
Spatial effects & Often negligible or well-controlled & Diffusion-limited gradients, biofilm structure \\
\hline
\end{tabular}
\end{table}

Traditional multiagent systems maintain agent populations with fixed or controllable size and exhibit deterministic or near-deterministic dynamics. Communication is typically digital, error-corrected, and operates on millisecond-to-second timescales. Biological consortia, by contrast, feature populations that grow and divide continuously (following logistic dynamics when resources are limited), exhibit intrinsically stochastic gene expression with substantial cell-to-cell variability, and communicate through analog diffusion of quorum sensing molecules over minute-to-hour timescales. The bandwidth limitations and spatial gradients in molecular communication fundamentally constrain the complexity of achievable coordination compared to electronic systems, while the absence of error correction mechanisms demands robust encoding of control signals.

Parameter uncertainty presents perhaps the most significant difference from traditional systems. Engineered multiagent systems typically operate with single-digit percentage uncertainty, well within the capabilities of standard robust control techniques. Biological systems exhibit substantially greater parameter variation across cells, growth conditions, and experimental contexts, arising from variable promoter strengths, fluctuating plasmid copy numbers, and context-dependent transcription rates. In our computational studies, robustness is typically assessed under parameter variations of 20\% or more \cite{fiore2017silico,martinelli2025multicellular}, and the control architectures must tolerate this level of uncertainty while maintaining stable regulation.

Unique to living systems are evolutionary dynamics that have no parallel in traditional multiagent control. Mutations that reduce metabolic burden by disabling controller genes confer immediate growth advantages: disabled controllers reduce cellular resource consumption, allowing faster division and competitive displacement of functional strains. This creates evolutionary pressure to eliminate precisely the control functions that maintain desired system behavior---a failure mode entirely absent from engineered systems where components do not compete for survival. This challenge becomes particularly significant in long-horizon applications such as continuous bioreactors or environmental deployment where many generations of growth occur over extended operation.

Despite these challenges, biological multiagent systems offer unique capabilities unavailable in traditional implementations: autonomous operation without external infrastructure, self-repair through population turnover, the ability to function in unstructured environments, and potential for adaptation through evolution when properly harnessed. Understanding these fundamental differences enables the design of control strategies that work with rather than against the constraints of living matter, translating classical multiagent control theory into forms appropriate for the stochastic, growing, evolving systems that microbial consortia represent.

\section{Validation of Multicellular Control: From Simulation to Experiment}
\label{sec:validation}

Translating the two-population control architecture from mathematical framework to functional biological system requires systematic validation through computational modeling followed by experimental implementation. This workflow enables rapid design iterations \textit{in silico} before investing in time-consuming wet-lab experiments.

\begin{figure}[tbp]
\centering
\includegraphics[width=0.85\linewidth]{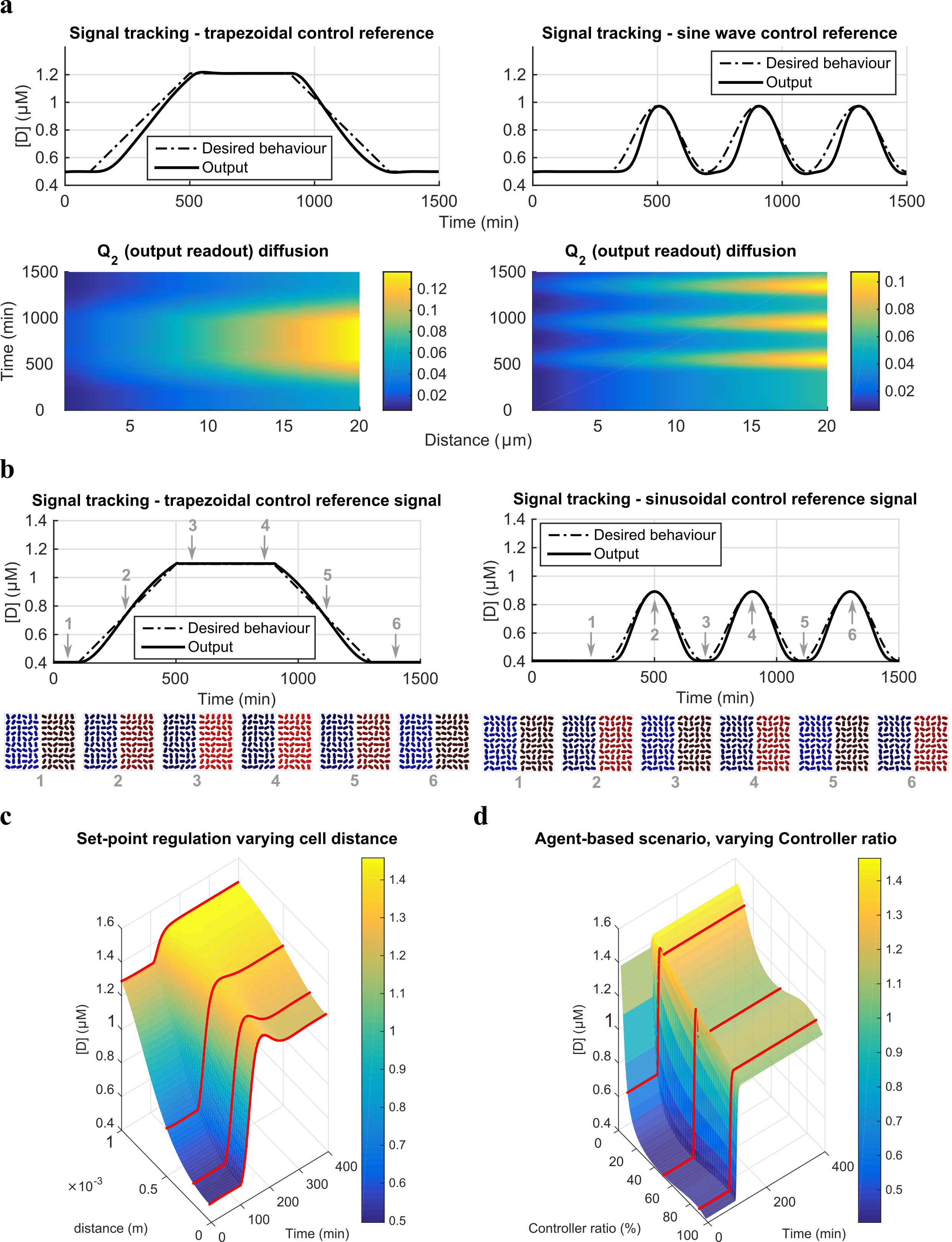}
\caption{\footnotesize{\bf \textit{In Silico} validation of the multicellular feedback control strategy using aggregate and agent-based models.}
(a) Aggregate population model: output of the Target population under set-point and time-varying reference signals, showing accurate regulation, limited overshoot, and stable steady-state behavior despite nonlinear dynamics and diffusive communication.
(b) Agent-based simulations in a microfluidic-like spatial domain: population-averaged Target output tracking trapezoidal and sinusoidal reference signals with negligible phase delay and low cell-to-cell variability.
(c) Robustness analysis: effect of increasing spatial separation between Controller and Target populations and of parameter perturbations ($\pm 20\%$) on regulation performance, showing preserved stability and reduced but nonvanishing dynamic range under severe communication attenuation.
(d) Composition independence and heterogeneity: regulation performance under varying Controller-to-Target population ratios and single-cell parameter variability, demonstrating that accurate control is maintained even when Controllers represent a small fraction of the consortium and when biological parameters vary across cells.
Together, the results indicate that the proposed distributed feedback architecture provides reliable regulation across spatial scales, biological noise, and population composition. All panels are reproduced from \cite{fiore2017silico}.}
\label{fig:computational_validation}
\end{figure}

\subsection{In Silico Validation}
\label{sec:in_silico}

The mathematical framework introduced in Section~\ref{sec:math_model} enables systematic validation of multicellular feedback control through both aggregate population models and agent-based simulations \cite{fiore2017silico,gorochowski2012bsim,matyjaszkiewicz2017bsim}. Together, these complementary approaches allow rapid exploration of design trade-offs under idealized conditions, followed by realistic assessment of spatial effects, stochastic gene expression, and single-cell heterogeneity prior to experimental implementation.

\subsubsection{Aggregate Population Models}

As shown in Section~\ref{sec:math_model}, aggregate models such as 
Equations~\eqref{eq:Z1}--\eqref{eq:Qx_ext} treat each population as homogeneous, 
enabling efficient simulation and systematic exploration of the design space. 
This formulation supports extensive parameter sweeps over promoter strengths, 
Hill coefficients, degradation rates, and QS production rates, as well as 
linearisation around equilibrium points to assess stability margins.

As shown in Fig.~\ref{fig:computational_validation}a, numerical simulations of Equations~\eqref{eq:Z1}--\eqref{eq:Qx_ext} confirm that the multicellular architecture introduced in Sec. \ref{sec:two_pops} achieves accurate tracking of different reference signals with negligible steady-state error, limited overshoot, moderate phase lag, 
and settling times on the order of 3--5 hours \cite{fiore2017silico}, which is compatible with bacterial growth dynamics under standard conditions. 
The aggregate framework also enables assessment of robustness through Monte Carlo simulations with random parameter perturbations, typically within $\pm 20\%$ of nominal values. Regulation performance remains satisfactory when parameters in both populations are simultaneously perturbed, and degrades only moderately when uncertainty is confined to the target population \cite{fiore2017silico}. Importantly, variations in target dynamics have limited impact on closed-loop behavior, supporting the notion of functional modularity as a single controller design can regulate targets with different intrinsic parameters without retuning.

To evaluate the impact of spatial separation and diffusion-limited signalling, 
we varied the distance between controller and target populations. As shown in 
Fig.~\ref{fig:computational_validation}c, increasing this distance attenuates 
both the effective control input and the measured output, reducing the achievable 
dynamic range. Nevertheless, closed-loop stability and convergence to the desired 
operating point are preserved over distances exceeding those typically encountered 
in microfluidic or colony-scale experimental platforms.

\subsubsection{Agent-Based Simulations}

While aggregate models enable efficient exploration of system-level behaviour, 
they neglect key biological features such as stochastic gene expression, cell 
growth and division, spatial crowding, and heterogeneous cell--cell interactions. 
Agent-based simulations using frameworks such as BSim~\cite{gorochowski2012bsim,matyjaszkiewicz2017bsim} 
address these limitations by explicitly modelling individual cells with stochastic 
intracellular dynamics, physical interactions, cell growth, motility, and spatially 
resolved QS diffusion fields (see Box~1 for details). Each cell evolves according 
to the same regulatory equations used in the aggregate model, but with parameters 
drawn from realistic distributions to capture cell-to-cell variability. 
Communication occurs through discretised reaction-diffusion processes in realistic 
microfluidic geometries.

Agent-based results (Fig.~\ref{fig:computational_validation}b) confirm that multicellular feedback regulation remains effective under biologically plausible spatial and stochastic conditions. Tracking of dynamic reference signals remains highly accurate, with population-averaged output closely following trapezoidal and sinusoidal trajectories. Convergence times remain comparable to aggregate model predictions, albeit with modest increases in transient overshoot due to the discrete nature of cell-to-cell communication. 

The agent-based framework further reveals how performance depends on population density and composition (Fig.~\ref{fig:computational_validation}d). Regulation remains effective across a wide range of total densities and controller-to-target ratios, including cases in which controllers represent only a small fraction of the consortium. Significant degradation emerges only at very low densities, where diffusion-mediated communication becomes insufficient to sustain coordination.

Together, the aggregate and agent-based validation studies provide strong computational evidence that multicellular feedback architectures can achieve reliable regulation and tracking across spatial scales, stochastic fluctuations, and population variability. Once this proof-of-concept is established, the abstract architecture presented in Fig.~\ref{fig:multicellular_control_framework}b can serve as a blueprint for biological implementation using appropriate biomolecular parts, as we describe in the following section.

\subsection{Experimental Implementation and Results}
\label{sec:experimental}

\begin{figure}[tbp]
\includegraphics[width=1\linewidth]{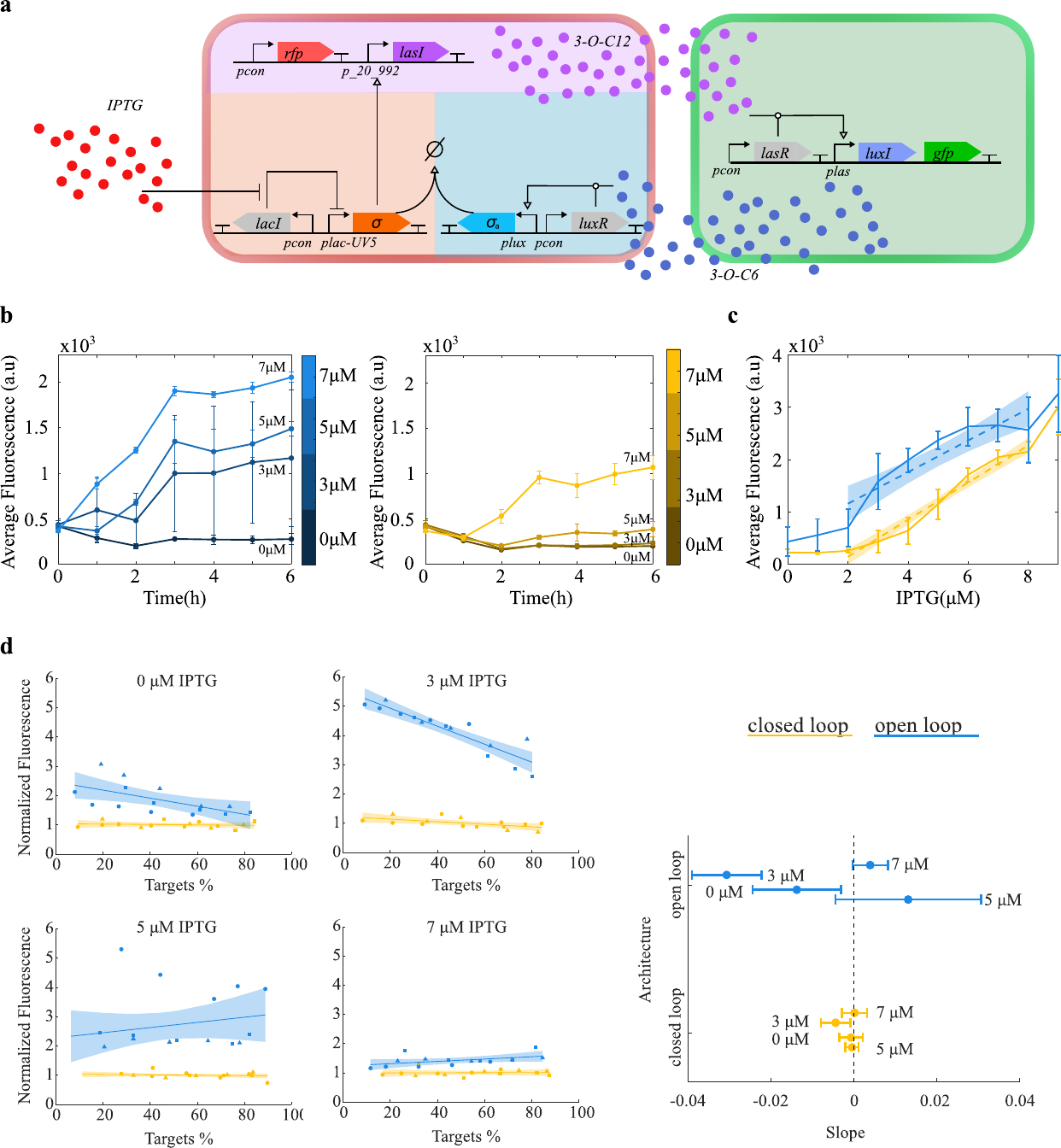}
\caption{\footnotesize\textbf{Validation of two-population feedback control.} 
(A) Schematic of two-population architecture with bidirectional QS communication and molecular titration for error computation. 
(B) Coefficient of variation across biological replicates for closed-loop (blue) versus open-loop (orange) control at different IPTG concentrations, showing sixfold reduction at 3~$\mu$M. 
(C) Normalized target fluorescence versus consortium composition (target percentage), demonstrating composition-independent output for closed-loop (flat, p$>$0.05) versus composition-dependent open-loop (negative slope, p$<$0.01). 
(D) Input-output relationships showing improved linearity for closed-loop ($R^2=0.91$) versus open-loop ($R^2=0.67$) control. 
All panels are reproduced from \cite{salzano2025vivo}.}
\label{fig:performance}
\end{figure}

The computational validation presented in the previous section established the theoretical feasibility of multicellular feedback control under idealized and realistic conditions. Translating these predictions into functional biological systems requires addressing implementation challenges that are difficult to capture fully in silico, including promoter leakiness, context-dependent gene expression, and the difficulty of predicting closed-loop behavior from component-level characterization (see Box~2 for a discussion of the biological implementation pipeline).

The first successful experimental demonstration of multicellular feedback control employed two engineered \emph{E.~coli} populations implementing distributed regulation through molecular titration mechanisms \cite{salzano2025vivo}. As illustrated in Fig.~\ref{fig:performance}a, controller cells compute the regulation error by comparing a reference signal provided via an external inducer (IPTG) with a feedback signal produced by target cells in the form of the QS molecule 3-O-C6-HSL. 
Error computation is implemented through a module first presented in \cite{annunziata2017orthogonal} which is based on molecular titration, a mechanism in which two molecular species bind and mutually sequester one another, effectively computing the difference between their concentrations. Specifically, the architecture employs $\sigma$/anti-$\sigma$ factor pairs \cite{rhodius2013design}: $\sigma$ factors are bacterial transcription initiation proteins, and anti-$\sigma$ factors are cognate inhibitors that bind and inactivate them. The reference signal induces $\sigma$ factor production while the feedback signal induces anti-$\sigma$ production; since these species sequester each other in a one-to-one stoichiometry, their relative abundances determine the effective amount of $\sigma$ available to activate downstream control genes, thereby encoding the magnitude of the error signal.

The resulting control signal is transmitted via a second quorum sensing channel (3-O-C12-HSL) to the target population, which modulates GFP expression accordingly while co-producing the feedback molecule 3-O-C6-HSL, thereby closing the loop. This design instantiates the mathematical framework introduced in Section~\ref{sec:math_model} using well-characterized genetic parts and orthogonal Lux/Las quorum sensing systems.

Figure~\ref{fig:performance}b--d summarises the experimental comparison between 
closed-loop feedback control and open-loop configurations in which controllers 
do not receive information about target output. Closed-loop operation substantially 
improved regulation reliability, reducing variability across biological replicates 
by approximately sixfold at intermediate reference levels (Fig.~\ref{fig:performance}b). 
Input--output linearity also increased markedly under feedback (Fig.~\ref{fig:performance}c). 
The coefficient of determination improves from $R^2 = 0.67$ in open loop to 
$R^2 = 0.91$ under closed-loop control, yielding predictable dose-response behaviour 
in which the output varies proportionally with the reference input. This property 
is essential for practical deployment where precise titration of the controlled 
variable is required. Output levels were largely independent of controller-to-target 
population ratios over the range 1:1 to 15:1 (Fig.~\ref{fig:performance}d), 
demonstrating the composition-independent regulation predicted by the mathematical 
analysis. Increases in controller abundance raise actuation, which elevates target 
output and hence feedback signal, suppressing further controller activity and 
stabilising the collective response despite compositional fluctuations. Settling 
times remained on the order of three hours for both configurations. Although maximum 
induction levels were lower under feedback than in open loop, the improved precision 
and robustness are more relevant for applications requiring consistent and reliable 
performance.

Beyond validating theoretical predictions, the experiments revealed implementation-level constraints not fully captured by computational models. Promoter leakiness limited the achievable dynamic range at low induction levels and required the use of additional operator sites to sharpen regulatory responses. Separately, the inclusion of degradation tags on key molecular species proved critical for two reasons: accelerating the dynamics to enable responsive feedback on timescales compatible with bacterial growth, and reducing sensitivity to population imbalances by preventing accumulation of signaling molecules. Furthermore, component-level characterization using proxy fluorescent reporters did not quantitatively predict closed-loop behavior, underscoring the necessity of testing complete feedback systems rather than extrapolating from isolated parts.

\subsection{Synthesis and Practical Design Insights}
The combined computational and experimental results highlight several design principles that are central to the feasibility of multicellular feedback control. First, physical separation between controller and target populations effectively reduces metabolic burden and mitigates retroactivity, enabling regulatory architectures whose complexity would be difficult to sustain within single cells. This separation allows controller populations to implement sensing and computation without directly perturbing the regulated biochemical processes, thereby preserving both stability and modularity. Second, orthogonal quorum sensing systems provide reliable bidirectional communication channels between populations. When appropriately selected and tuned, Lux- and Las-based signaling exhibits sufficiently low crosstalk to support independent feedback pathways, even in dense cultures. Third, molecular mechanisms such as $\sigma$/anti-$\sigma$ factor titration offer practical biochemical realizations of abstract control operations, including error computation, allowing classical feedback concepts to be mapped onto genetic circuit implementations. Finally, population-level redundancy introduces an inherent form of fault tolerance: feedback regulation maintains stable collective behavior despite substantial stochastic variability at the single-cell level.

At the same time, validation studies highlight limitations that remain critical for deployment beyond controlled laboratory settings. In large batch cultures or applications requiring extended autonomous operation, maintenance of stable population ratios may become challenging, as even moderate differences in metabolic burden can lead to compositional drift through competitive exclusion; bioreactor environments with continuous dilution are less susceptible to this issue, but long-term batch processes or environmental applications may require additional safeguards. Reactor-scale operation introduces further constraints, since mixing times and transport delays may limit the effective bandwidth of quorum-sensing communication, particularly in large-volume or spatially structured environments. Evolutionary stability also represents a persistent challenge, as mutations that disable controller functions can confer growth advantages, generating selection pressures that gradually erode regulatory performance over many generations. Moreover, spatial organization in biofilms or poorly mixed systems creates concentration gradients that weaken coupling between populations, suggesting the need for either enhanced mixing, increased signaling capacity, or control architectures that explicitly account for spatial dynamics.

Despite these challenges, the qualitative agreement between computational predictions and experimental observations supports the view that multicellular feedback control is not merely a theoretical construct but a practically realizable engineering strategy. The modeling framework captures the key mechanisms underlying composition-independent regulation and provides a useful tool for guiding circuit design, while experimental validation clarifies the operational regimes in which distributed control can be reliably implemented. Together, these results provide a foundation for extending multicellular control architectures toward more complex controllers and toward application domains where robustness, adaptability, and long-term stability are essential requirements.

\section{Advanced Controllers: The Multicellular PID Family}
\label{sec:pid}

Building on the basic two-population feedback architecture, subsequent studies have demonstrated that increasingly sophisticated biomolecular control structures inspired by classical proportional-integral-derivative (PID) controllers can be implemented in a modular multicellular fashion. While single-cell integral feedback has been demonstrated through antithetic control mechanisms (guaranteeing robust perfect adaptation) \cite{aoki2019universal} and an embedded PID controller was analyzed in-silico in \cite{chevalier2019design}, the multicellular approach offers complementary advantages. Namely, each control action can be realized by a functionally distinct population, distributing the genetic burden and enabling independent tuning of controller parameters, potentially enabling plug-and-play synthesis of different controllers in the PID family \cite{martinelli2025multicellular,filo2022hierarchy}.

The resulting architecture, illustrated in Fig.~\ref{fig:multicellular_control_framework}c, assigns proportional, integral, and derivative actions to separate cell populations. Each population receives the error signal (encoded as a quorum-sensing molecule) and produces a contribution to the overall control input corresponding to its designated action. These contributions combine in the shared extracellular medium before reaching the target population, effectively providing the equivalent biomolecular P, I, and D terms found in the classical parallel PID structure. The key distinction from classic PID implementations lies in how each action is realized biomolecularly. Briefly, proportional action arises from direct transcriptional activation, integral action from the accumulating imbalance in an antithetic motif, and derivative action from an incoherent feedforward loop that responds to the rate of change of its input. It is important to note that these biomolecular implementations are inherently nonlinear, governed by Hill-type response functions and saturation effects, and resemble their classical linear counterparts only within certain operating regions where the underlying biochemical reactions approximate the desired input-output relationships. A detailed analysis of these operating regimes and the conditions under which the analogy to classical PID control holds can be found in  \cite{martinelli2022multicellular,filo2022hierarchy,martinelli2025multicellular}. Despite these nonlinearities, the distributed realization enables systematic controller design while preserving modularity and reuse of controller populations.

The foundational multicellular controller implements integral action through an antithetic motif distributed across two populations, as described in Section~\ref{sec:two_pops}. Multicellular PI control \cite{martinelli2022multicellular} extends this by adding a proportional branch to improve transient performance. While integral action alone can exhibit slow convergence and significant overshoot, the proportional term accelerates the response and provides additional damping. Theoretical analysis further shows that proportional-integral architectures can reduce output variability arising from stochastic gene expression \cite{briat2020silico}. Numerical and agent-based validations confirm that PI architectures achieve robust set-point regulation even under significant parameter variability, stochastic gene expression, and population heterogeneity, as well as under variations in controller-to-target population ratios \cite{martinelli2022multicellular}.

Multicellular PD control \cite{martinelli2023multicellular} is primarily aimed at improving transient response. By introducing a derivative branch that responds to the rate of change of the error signal, the closed-loop system exhibits increased damping and reduced overshoot over broad regions of the gain space. Agent-based simulations confirm that PD architectures can significantly attenuate oscillatory transients and shorten settling times compared to proportional control, while maintaining comparable steady-state accuracy. However, as in classical control, derivative action alone does not remove steady-state error and remains sensitive to noise, limiting its standalone applicability for precise regulation in biological systems.

A unified analysis of the multicellular P, PI, PD, and PID families shows that the benefits of integral and derivative actions are largely complementary \cite{martinelli2025multicellular}. Integral action primarily determines robustness of steady-state regulation to biological uncertainty and parameter dispersion, whereas derivative action expands the region of stabilizing controller gains and improves transient shaping. In combined PID architectures, agent-based simulations indicate improved tolerance to cell-to-cell variability and faster convergence compared to PI control alone, while preserving near-zero steady-state error. Across all architectures, regulation remains effective under realistic communication constraints, including diffusion-limited signaling and spatial separation between populations, as well as moderate heterogeneity in cellular parameters, confirming that feedback compensation is not compromised by the physical distribution of control functions.

These results support the view that multicellular PID controllers constitute a family of modular architectures rather than a single fixed design. Depending on application requirements, designers may prioritize robustness of steady-state regulation (favoring PI or PID structures) or improved transient performance and damping (favoring PD or PID), while retaining the advantages of distributed implementation, reduced per-cell genetic burden, and composition independence that motivate multicellular control strategies. However, increasing the number of controller populations introduces additional challenges: maintaining stable coexistence among three or four distinct populations becomes progressively more difficult as differential growth rates accumulate. The choice of controller complexity must therefore balance the benefits of improved dynamic performance against the practical difficulty of sustaining multi-population consortia over extended operation, a problem we address in the following section.

\begin{figure}[btp]
\includegraphics[width=1\linewidth]{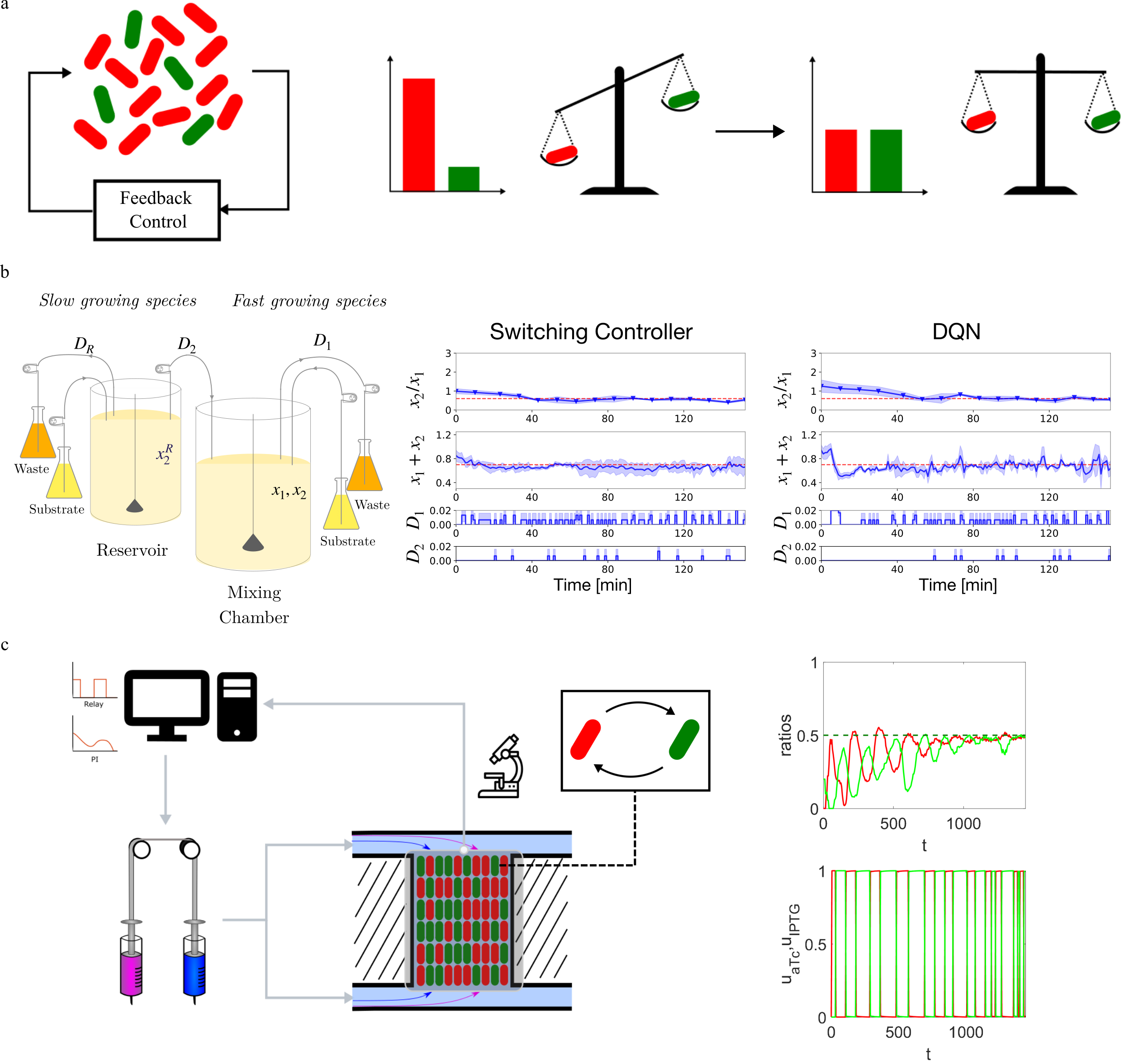}
\caption{\footnotesize \textbf{Approaches to ratiometric control.} 
(A) Schematic of a composition control scheme (reproduced from \cite{salzano2022ratiometric}). 
(B) Representation of a dual chamber bioreactor architecture for composition control (left), and experimental data of biomass and composition regulation with a switching and learning-based controller (reproduced from \cite{brancato2025bioreactor}).
(C) Composition control in microfluidics exploiting a reversible memory machanism. On the right panel the microfluidics platform is represented. On the right panel a representative example of regulation with a switching controller is shown (reproduced from \cite{salzano2022ratiometric}). }
\label{fig:composition_control}
\end{figure}

\section{Population Composition Control}
\label{sec:composition}

Population composition control (also referred to as ratiometric control) is a fundamental challenge unique to multicellular systems. It concerns the maintenance of stable coexistence of functionally distinct cellular subpopulations in the presence of differential growth rates arising from unequal metabolic costs, resource competition, and environmental variability \cite{qian2017resourcecompetition,ceroni2015burden}.

Because engineered functions typically impose unequal metabolic loads, even modest growth-rate mismatches can generate sustained drift in population ratios over experimentally relevant timescales. From a control perspective, composition control therefore constitutes an additional control layer acting on population-level states. This layer is coupled to, but conceptually distinct from, the design of intracellular and intercellular feedback dynamics discussed in previous sections, and determines whether the assumptions under which molecular controllers are designed remain valid over time.

It is useful to distinguish two orthogonal design dimensions in population 
composition control. The first concerns the \emph{locus of feedback implementation}, 
which can be external (reactor- or computer-mediated orchestration) or embedded 
within genetic circuits. The second concerns the \emph{controlled variable}, which 
may be the relative abundance of distinct strains or the fraction of cells occupying 
different phenotypic states within a single strain. 
Figure~\ref{fig:composition_control} illustrates three representative approaches 
within this design space: external feedback regulation of strain ratios, ratiometric 
control via reversible phenotypic switching, and reactor-level orchestration using 
dual-chamber architectures.

\subsection{External orchestration of population ratios}

One class of approaches treats consortium composition as an externally regulated process variable, manipulated through environmental or reactor-level inputs. In a single chemostat, the dilution rate can act as a control input to regulate the ratio of two populations when their growth-rate curves intersect, enabling selective advantage across different operating regions. In this setting, feedback strategies based on gain scheduling or switching control have been shown to stabilize coexistence and regulate population ratios without requiring genetic modification of the strains \cite{Fiore2021Chemostat,lee2025directing,gutierrez2022dynamic,bertaux2022enhancing}. This formulation provides a clear control-theoretic interpretation of coexistence as a reachable operating point under suitable actuation (Fig.~\ref{fig:composition_control}A). 

This perspective also reveals a fundamental limitation. When species are non-complementary (meaning that one population outgrows the other across the relevant operating range), the competitive exclusion principle implies that coexistence cannot be guaranteed at steady state using dilution-rate actuation alone \cite{Balagadde2008},\cite{cardinale2012context}. To overcome this structural limitation, dual-chamber bioreactor platforms have been proposed \cite{Brancato2024DualChamber}, in which the slower-growing strain is cultivated in a reservoir and periodically injected into a mixing chamber hosting the co-culture (Fig.~\ref{fig:composition_control}B).  This architecture introduces an independent actuation channel through the inter-chamber transfer rate, allowing population ratios to be regulated independently of in-chamber growth competition. Model-based analyses show that switching feedback strategies can robustly regulate both total biomass and composition for strain pairs that would otherwise be incompatible in a single-reactor setting \cite{Brancato2024DualChamber}. Subsequent work has extended this framework toward experimentally grounded and data-efficient control, combining model-based and learning-based controllers trained via sim-to-real pipelines, with \emph{in vivo} validation demonstrating reference tracking, recovery from perturbations, and robustness to parameter uncertainty \cite{Brancato2024Learning}. These results illustrate how reactor-level orchestration can provide a stable experimental substrate for testing multicellular molecular controllers under controlled population ratios.

Related external orchestration strategies have been developed within cybergenetic control frameworks, where optical or environmental actuation is combined with real-time feedback. Optogenetic regulation of differentiation has been used to dynamically control the composition of yeast consortia, enabling reversible and tunable allocation of cells to distinct functional roles under light-driven feedback \cite{aditya2021light}. Other strategies employ externally driven switching inputs to toggle genetic circuits associated with different phenotypes in a consortium (Fig.~\ref{fig:composition_control}C) \cite{salzano2022ratiometric}. While these approaches regulate composition through phenotypic switching rather than direct strain competition, they share the ratiometric objective of controlling functional allocation. Similarly, optogenetic modulation of antibiotic resistance has been used to regulate effective growth rates and enforce desired population ratios in bacterial co-cultures under closed-loop control \cite{gutierrez2022dynamic}. These approaches typically assume complementary strains, but demonstrate how external feedback can compensate for biological variability and enforce coordination objectives that are difficult to achieve through embedded circuits alone.

\subsection{Embedded composition control}

In embedded composition control, feedback regulation of population structure is implemented directly within genetic circuits, without requiring continuous external intervention. A prominent mechanism in this class is phenotypic switching, whereby functional roles are regulated within a genetically homogeneous population through reversible transitions between cellular states \cite{weill2019optimal}. In this formulation, the control objective is the regulation of the fraction of cells occupying each phenotype rather than the abundance of physically distinct strains.

Other studies have demonstrated fully embedded mechanisms in which population composition emerges from genetic feedback alone. A notable example is provided by differentiation circuits that couple cellular state to growth or metabolic capacity. In a recent experimental study, a multi-stage differentiation program in \textit{E.~coli} was shown to maintain a stable mixture of growth and production phenotypes while suppressing takeover by fast-growing mutants, thereby enhancing evolutionary stability \cite{Glass2024}. In this setting, differentiation effectively implements a population-level negative feedback mechanism that regulates functional heterogeneity and mitigates competitive exclusion.

Theoretical work has similarly shown that state-dependent phenotypic switching can stabilize coexistence even when static ecological equilibria predict dominance of a single phenotype. Models incorporating responsive switching dynamics predict robust maintenance of mixed populations and increased resilience to perturbations, suggesting that controllable switching can substitute for explicit population-level actuation \cite{Haas2022}. Related analyses of bistable and co-repressive circuits further indicate that embedded multistability and delayed feedback can generate stable ratios or oscillatory population structures in microbial consortia \cite{Sadeghpour2017}.

Another class of embedded strategies relies on self-limiting growth implemented 
through quorum-sensing--regulated killing or growth inhibition. Early work 
demonstrated that density-dependent toxin expression can regulate total population 
size through programmed cell death~\cite{You2004}. This concept was later extended 
to multi-strain systems using kill--rescue or predator--prey architectures, in 
which reciprocal signalling modulates survival and leads to stable coexistence or 
sustained oscillations~\cite{Balagadde2008,chen2015emergent}. More recently, 
orthogonal quorum-sensing--controlled lysis circuits have been used to stabilise 
otherwise competitive co-cultures by selectively penalising fast-growing populations, 
enabling long-term coexistence without enforced metabolic 
dependencies~\cite{Scott2017}.

A closely related but conceptually distinct line of work focuses on embedded 
regulation of population growth rates through genetic feedback, providing a 
foundational actuation mechanism for composition control. Fusco et 
al.~\cite{Fusco2022Embedded} proposed an architecture in which cells self-regulate 
their growth rate using a tunable expression system coupled with quorum sensing. 
By modulating the production of a growth-inhibitor protein as a function of 
population density, the circuit implements a fully embedded negative feedback loop 
that stabilises biomass density and allows the reference set-point to be adjusted 
through genetic or slowly varying external tuning. Although demonstrated in a 
single-strain setting, this approach highlights how embedded regulation of effective 
growth rates can serve as an inner control layer to support long-term coexistence 
and robustness in multicellular consortia, either as a standalone density controller 
or in combination with higher-level composition or differentiation mechanisms.

From a control-theoretic standpoint, embedded composition regulation can be interpreted as feedback acting on population growth, death, or switching probabilities rather than directly on molecular outputs. Formal architectures have been proposed in which multiple feedback loops regulate both total population size and relative abundance through differentiation, death, or phenotypic transitions \cite{ren2017population}. These formulations highlight strong analogies with coordination control in multi-agent systems, with actuation channels intrinsically tied to cellular physiology.

Overall, embedded strategies based on phenotypic switching and self-regulation provide a complementary route to composition control that reduces reliance on external actuation and reactor-level intervention. While these approaches shift design complexity toward genetic circuit stability and evolutionary robustness, they offer the prospect of long-term operation and tighter integration between molecular objectives and population-level organization. As such, they complement external orchestration and multi-strain architectural approaches within the broader design space of multicellular feedback control.

\section{Conclusions and Future Directions in Multicellular Control}
\label{sec:conclusions}
Multicellular feedback control in synthetic microbial consortia provides a principled way to distribute sensing, computation, and actuation across interacting populations, overcoming key limitations of single-cell control strategies related to metabolic burden, retroactivity, and lack of modularity. The results reviewed in this paper demonstrate that feedback regulation can be achieved robustly despite stochastic gene expression, spatial communication constraints, and heterogeneous population compositions, establishing multicellular control as a viable engineering paradigm rather than a purely conceptual proposal.

At the modeling level, an important open challenge is bridging the gap between current agent-based simulations and the physical environments relevant for large-scale cultivation and field deployment. While existing models capture stochastic intracellular dynamics and diffusion-mediated communication, they typically neglect realistic transport phenomena such as advection, mixing heterogeneity, and shear effects. Developing reduced-order or multiscale models that integrate population dynamics with representations of fluid transport will be essential for predicting performance and stability of multicellular controllers beyond microfluidic and small-batch settings.

Experimental infrastructure also remains a limiting factor. Most bioreactors are designed for monocultures and provide limited support for independent monitoring and actuation of multiple interacting populations. Scaling multicellular control to liter-scale or larger systems introduces fundamental constraints on signaling bandwidth and response times, as well as increased risks of contamination and population imbalance. Reactor architectures and sensing technologies specifically tailored to consortia, together with automated control of dilution and feeding, will be critical to move multicellular feedback strategies from laboratory demonstrations toward sustained long-term operation.

From a control design perspective, current multicellular controllers largely 
implement low-order architectures inspired by classical PID control, though only 
the antithetic integral motif has been validated experimentally to date; the 
proportional and derivative extensions remain computationally demonstrated but 
await \emph{in vivo} implementation. Extending these concepts to incorporate 
adaptive, stochastic, or predictive control mechanisms could substantially improve 
performance under uncertain and time-varying growth conditions. However, realising 
such strategies requires solving the biological realizability problem of identifying 
molecular mechanisms and communication patterns capable of implementing abstract 
control laws under severe constraints on reaction kinetics, noise, and resource 
availability. Communication delays, saturation effects, and crosstalk further 
impose performance limits that must be explicitly accounted for in controller 
synthesis.

A distinctive and unresolved aspect of multicellular control is long-term ecological and evolutionary stability. Differences in growth rates, metabolic costs, and mutation-driven loss of function can destabilize carefully tuned architectures over extended time horizons. Achieving persistent regulation may require integrating feedback control with ecological design principles, including mutualistic dependencies, density-dependent growth regulation, or spatial structuring that aligns evolutionary fitness with control objectives. Developing control architectures that remain functional under evolutionary pressure represents a fundamental challenge that has no direct analogue in traditional engineered systems.

Despite these challenges, the range of potential applications for multicellular control is broad, spanning biomanufacturing, environmental remediation, and therapeutic interventions \cite{hwang2020engineering,arif2020plant,Shong2012}. In all these domains, the ability to coordinate specialized populations through feedback offers opportunities to achieve levels of robustness and functional complexity that are difficult to attain within single-cell designs. Continued progress will depend on tighter integration between control-theoretic analysis, circuit-level synthetic biology, and scalable experimental platforms designed explicitly for consortium-based operation.

In summary, multicellular control shifts the focus of biological regulation from isolated engineered cells to coordinated populations acting as distributed control systems. By treating microbial consortia as controllable multiagent systems, this paradigm opens new directions for both control theory and synthetic biology, while posing uniquely interdisciplinary challenges that will require sustained collaboration across traditionally separate research communities.

\section*{Acknowledgements}

The research program on multicellular feedback control described in this review represents the outcome of a decade-long collaboration between the University of Naples Federico II in Italy and the University of Bristol in the U.K. I am deeply grateful to my colleagues Nigel Savery (School of Biochemistry, Bristol), Claire Grierson (School of Biological Sciences, Bristol), and Lucia Marucci (Department of Engineering Mathematics, Bristol) for their sustained partnership throughout this endeavor. The work would not have been possible without the dedicated contributions of the PhD students and postdoctoral researchers in Bristol and Naples who tackled different aspects of the project over the years: Thomas Gorochowski, Gianfranco Fiore, Fabio Annunziata, Antoni Matyjaszkiewicz, Criseida Zamora-Chimal, Barbara Shannon, Davide Fiore, Davide Salzano, Sara Brancato and Vittoria Martinelli. This research was supported over the years by funding from the UK Biotechnology and Biological Sciences Research Council (BBSRC), the UK Engineering and Physical Sciences Research Council (EPSRC), the Italian Ministry of University and Research (MUR), and the Bristol Centre for Synthetic Biology (BrisSynBio).

The author wishes to thank Dr.\ Davide Salzano (University of Naples Federico II, Italy) for his thoughtful contributions to the boxed sidebars, for meticulous proofreading of the manuscript, and for providing useful comments. The use of AI-based writing assistance tools for language editing and stylistic refinement of portions of the text is also acknowledged. All technical content, interpretations, and conclusions remain the sole responsibility of the author.
\bibliographystyle{unsrt}
\bibliography{bibliography}

\begin{csmbox}{BSim Framework for Agent-Based Modeling}

\begin{center}
\includegraphics[width=\linewidth]{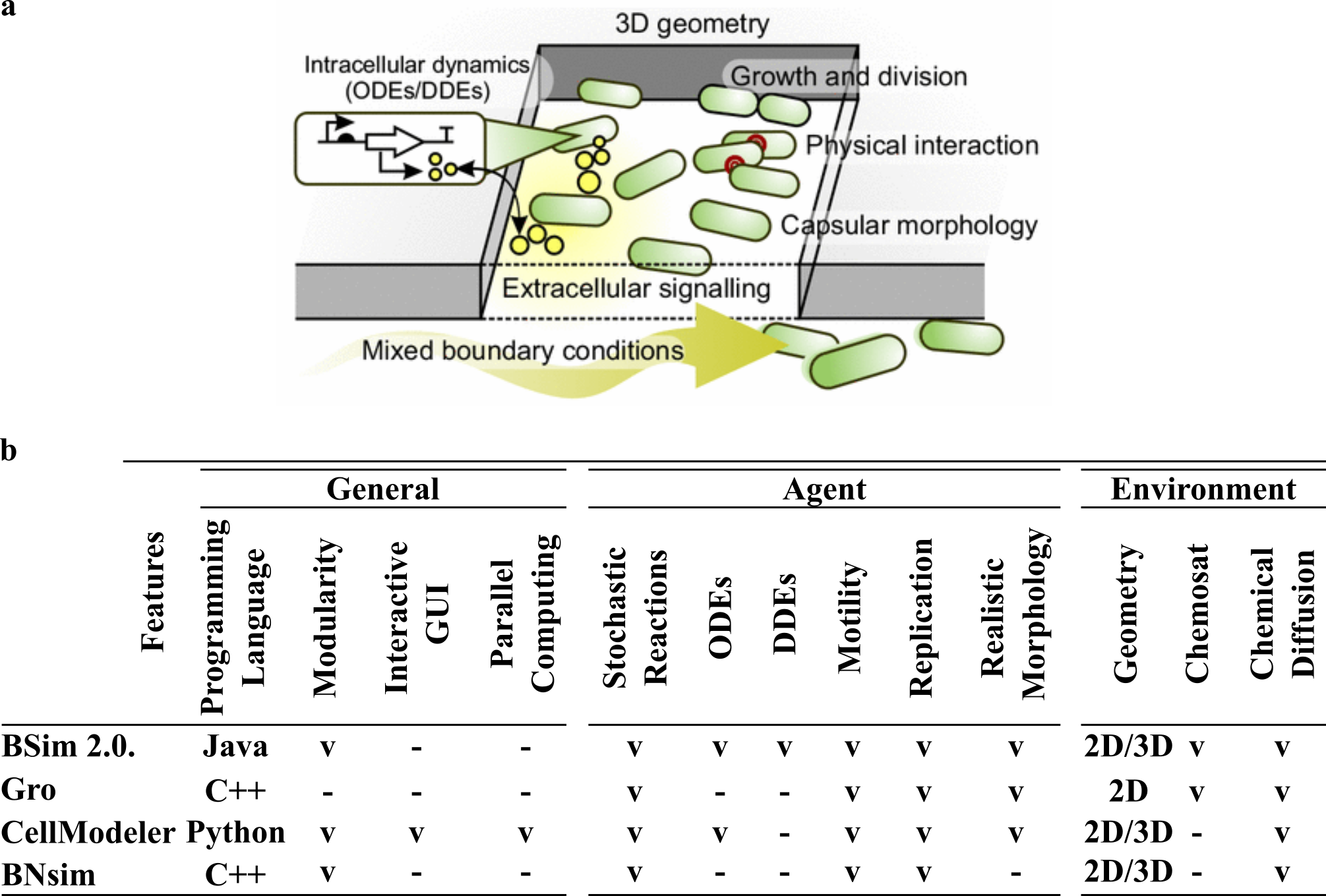} \footnotesize
\captionof{figure}{\textbf{a} Agent-based simulation explicitly models intracellular dynamics, together with cell growth, division, mechanical interactions, and extracellular signaling. \textbf{b} Comparison of representative agent-based simulators for microbial populations. Both panels are reproduced from \cite{matyjaszkiewicz2017bsim}.}
\label{fig:BSim_agent_based}
\end{center}

Accurate analysis of multicellular feedback controllers requires models that capture not only intracellular gene regulation but also the spatial and physical interactions among cells and their environment. Agent-based modeling addresses this need by representing each cell individually and explicitly simulating growth, division, movement, mechanical contact, and diffusion of signaling molecules. As illustrated in Fig.~\ref{fig:BSim_agent_based}, this approach naturally links single-cell dynamics to emergent population-level behaviors.

\textbf{BSim} is an open-source, Java-based framework for agent-based simulation of bacterial populations, first introduced in 2012 \cite{gorochowski2012bsim} and extended in version~2.0 \cite{matyjaszkiewicz2017bsim}. Each cell carries its own gene regulatory network, which can be described using ordinary, delayed, or stochastic differential equations. The simulator also models realistic geometries such as microfluidic chambers and chemostats, together with diffusion, degradation, and transport of extracellular species.

Compared with well-mixed population models, agent-based simulators offer two key advantages. First, they reveal emergent collective phenomena---including quorum sensing, spatial pattern formation, and coexistence dynamics—that depend on local interactions. Second, they capture cell-to-cell variability and rare stochastic events that are typically averaged out in deterministic descriptions.

In practice, such \emph{in silico} experiments provide a valuable design tool for multicellular control. Candidate architectures can be tested, tuned, and stress-tested under realistic variability before experimental implementation, reducing both development time and laboratory cost.

The latest release of BSim, together with documentation and example models, is available at \url{https://github.com/bsim-bristol/bsim}. Alternative packages compared with BSim in Figure \ref{fig:BSim_agent_based}b are described in \cite{gutierrez2017gro,cellmodeller,sneddon2011networkfree}.
\end{csmbox}S

\begin{csmbox}{Experimental Implementation Guide: Design and Characterization}

\begin{center}
\includegraphics[width=\linewidth]{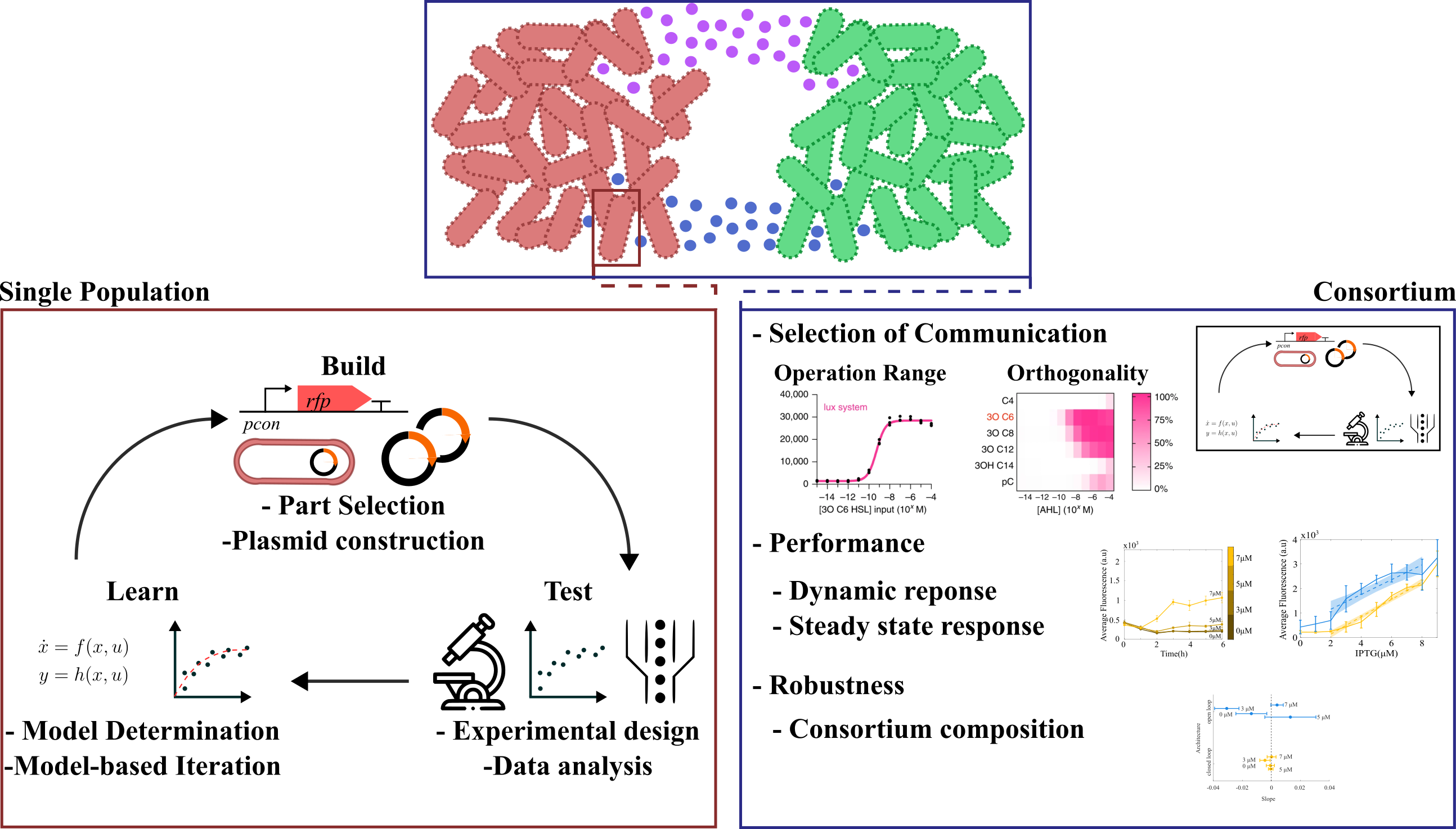}
\captionof{figure}{Pipeline for implementing a multicellular feedback control scheme \emph{in vivo}. Each strain is developed through a build–test–learn cycle before integration into the final consortium.}
\label{fig:experimental_pipeline}
\end{center}

Implementing a multicellular feedback controller requires the co-design of two interacting bacterial strains that play the roles of \emph{Controller} and \emph{Target}. As illustrated in Fig.~\ref{fig:experimental_pipeline}, development follows an iterative build–test–learn workflow applied first to each population individually and only later to the integrated consortium.

For each strain, suitable biological components must be selected to implement the desired control functionality. For example, the Controller requires a comparator module that measures the deviation between the Target’s gene expression and a reference signal. Such behavior can be realized using an annihilation (antithetic) motif, here implemented through $\sigma$/anti-$\sigma$ factor pairs \cite{annunziata2017orthogonal}. This motif provides a biomolecular analogue of integral action and enables robust regulation of gene expression.

Once the functional modules are identified, the gene network is partitioned across compatible plasmids. This design balances construct size, modularity, and genetic stability while respecting replication-origin constraints of the host chassis. Modular partitioning also facilitates independent testing and debugging of subsystems.

Each construct is assembled using standard molecular biology techniques \cite{thomas2015dna,engler2013golden} and experimentally characterized. Promoter strengths, ribosome binding sites, and degradation rates are tuned to match model predictions. Measured input–output responses (e.g., from batch cultures or flow cytometry) are used to update mathematical models and guide subsequent redesign steps, closing the build–test–learn loop \cite{litcofsky2012iterative}.

This model-informed workflow allows multicellular control architectures to be validated \emph{in silico} and experimentally at the single-strain level before moving to full consortium integration, significantly reducing development time and cost.

\end{csmbox}

\begin{csmbox}{Experimental Implementation Guide (cont.): Consortium Integration and Validation}

After individual characterization, the Controller and Target strains are integrated into a consortium that performs the desired collective task. At this stage, the primary design challenge becomes establishing reliable communication and validating closed-loop behavior at the population level.

Bidirectional information exchange is typically achieved through quorum-sensing molecules that diffuse across the medium. Because multiple signaling channels may introduce cross-talk, orthogonal acyl-homoserine lactone (AHL) pairs must be selected to minimize unintended coupling \cite{kylilis2018tools}. In our implementation, 3-O-C6 and 3-O-C12 HSL were chosen for their minimal interference and suitable operating ranges.

System-level validation follows a control-oriented protocol. Open-loop experiments first verify that the Controller population can influence Target gene expression. The feedback pathway is then enabled to close the loop, and the complete architecture is evaluated using classical control metrics, including set-point tracking, transient response, steady-state accuracy, and robustness to disturbances and parameter variability. These experiments establish whether the consortium behaves as a reliable multicellular feedback controller.

Several experimental platforms can be employed. Batch or chemostat cultures coupled with flow cytometry provide scalable and cost-effective measurements of population-averaged behavior \cite{bertaux2022enhancing}. In contrast, microfluidic devices combined with time-lapse microscopy enable continuous, single-cell monitoring and reveal spatial effects and heterogeneity \cite{pedone2019tunable}. While microfluidics offers higher temporal and spatial resolution, it requires more complex experimental setups and careful strategies to maintain long-term coexistence of multiple strains.

In practice, combining bulk and microfluidic experiments provides complementary insight into both population level performance and single-cell variability, enabling comprehensive validation of multicellular control strategies.

\end{csmbox}
\end{document}